\documentclass[fleqn,10pt]{SelfArx}
\usepackage{lipsum}
\usepackage{natbib}

\definecolor{color1}{RGB}{0,0,90} 
\definecolor{color2}{RGB}{0,20,20} 

\usepackage{hyperref} 
\hypersetup{hidelinks,colorlinks,breaklinks=true,urlcolor=color2,citecolor=color1,linkcolor=color1,bookmarksopen=false,pdftitle={Title},pdfauthor={Author}}

 \def\gsim{\lower.4ex\hbox{$\;\buildrel >\over{\scriptstyle\sim}\;$}}
 \def\lsim{\lower.4ex\hbox{$\;\buildrel <\over{\scriptstyle\sim}\;$}}

\JournalInfo{Solar-Terrestrial Physics, 2017. Vol. 3. Iss. 1, pp. 3--18} 
\Archive{}
\PaperTitle{SIBERIAN RADIOHELIOGRAPH: FIRST RESULTS}
\Authors{S.V. Lesovoi\textsuperscript{1}*, A.T.
Altyntsev\textsuperscript{1}, A.A. Kochanov\textsuperscript{1,2},
V.V. Grechnev\textsuperscript{1}, A.V. Gubin\textsuperscript{1},
D.A. Zhdanov\textsuperscript{1}, E.F. Ivanov\textsuperscript{1},
A.M. Uralov\textsuperscript{1}, L.K.
Kashapova\textsuperscript{1,2}, A.A. Kuznetsov\textsuperscript{1},
N.S. Meshalkina\textsuperscript{1}, R.A. Sych\textsuperscript{1}}

 \affiliation{\textsuperscript{1}\textit{Institute of Solar-Terrestrial Physics SB RAS, Irkutsk, Russia}}
 \affiliation{\textsuperscript{2}\textit{Irkutsk State University, Irkutsk, Russia}}
 \affiliation{*\textbf{E-mail}: svlesovoi@gmail.com}

\Keywords{Sun; Radio emission; Solar flares; Radio telescopes;
Negative bursts}
\newcommand{\keywordname}{Keywords}

\Abstract{Regular observations of active processes in the solar
atmosphere have been started using the first stage of the
multiwave Siberian Radioheliograph (SRH), a T-shaped 48-antenna
array with a 4--8 GHz operating frequency range and a 10 MHz
instantaneous receiving band. Antennas are mounted on the central
antenna posts of the Siberian Solar Radio Telescope. The maximum
baseline is 107.4 m, and the angular resolution is up to
$70^{\prime \prime}$. We present examples of observations of the
solar disk at different frequencies, ``negative'' bursts, and
solar flares. The sensitivity to compact sources reaches 0.01
solar flux units ($\approx 10^{-4}$ of the total solar flux) with
an accumulation time of about 0.3\,s. The high sensitivity of SRH
enables monitoring of solar activity and allows studying active
processes from characteristics of their microwave emission,
including faint events, which could not be detected previously.}

\begin{document}

\flushbottom 

\maketitle 


\thispagestyle{empty} 

\section{Introduction} 

The Institute of Solar-Terrestrial Physics of Siberian Branch of
the Russian Academy of Sciences has developed the largest complex
of instruments for ground-based observations of various space
weather phenomena and their sources -- from events occurring in
the solar atmosphere to ionospheric plasma disturbances. A
prominent place in this complex, as well as in the international
cooperation, belongs to the instrumentation for monitoring and
diagnostics of solar active processes by radio astronomy methods.
Information supplied by radio observations is significant for both
fundamental solar physics and practical use, especially in Russia
which still lacks continuous observations of solar activity from
space.

\begin{figure*} 
           \centerline{\includegraphics[width=\textwidth]
           {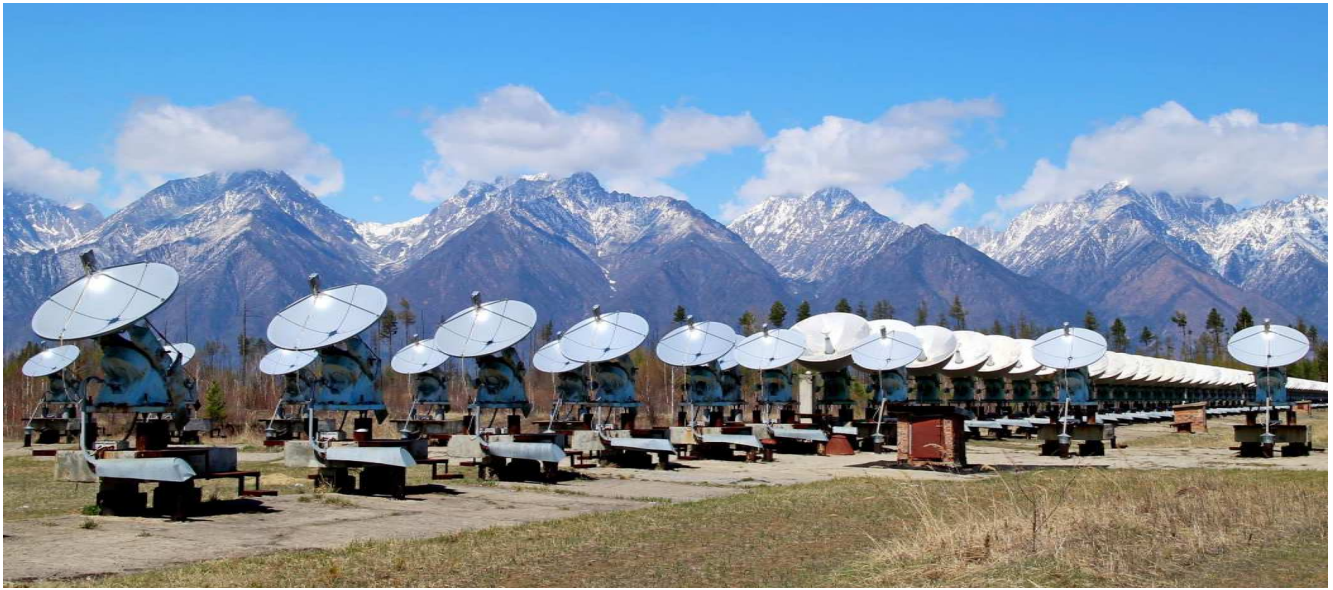}
           }
\caption{The central part of the SRH antenna array and the
northern arm of SSRT.}
 \label{fig:radioheliograph}
\end{figure*}

The radio telescopes are located in the Radioastrophysical
Observatory (RAO) (the Badary area, the Buryat Republic). Solar
observations are made during daylight hours from 00 to 10 UT in
summer and from 02 to 08 UT in winter. The observatory located
away from communities, the radio noise level is low. In RAO, a
multiwave Siberian Radioheliograph (SRH,
Figure~\ref{fig:radioheliograph}) is under construction.
Total-flux spectropolarimeters observe solar emission in a
frequency band from 50 MHz to 24 GHz. Data from the Badary
Broadband Microwave Spectropolarimeter (BBMS) in the 4--8 GHz
operating frequency band of the first stage of SRH are also
available \citep{ZhdanovZandanov2015}.

Observational data from RAO's radio telescopes are used by Russian
and foreign scientists for studies of fundamental problems in
solar physics, including those carried out under international
programs. Prospects for the use of the observational data to solve
applied problems determine the significant contribution of the
Fedorov Institute of Applied Geophysics, Roshydromet, into the
development of the radio telescope complex. Solar activity data,
including real-time ones, are available online at
\url{http://badary.iszf.irk.ru}.

Radio observations provide unique information on processes of
plasma heating and particle acceleration in the solar atmosphere.
The total solar flux at a frequency of 2.8 GHz (the so-called
10.7\,cm index) is the most objective assessment of the current
level of solar activity and its variability over several solar
cycles. The 10.7\,cm index is an input parameter in various models
of near-Earth space, magnetosphere and ionosphere.

The advantage of solar activity monitoring in the radio range is
its all-weather capability, as weather conditions have little
effect on a received signal. The cost of radio telescopes is
relatively low compared to space missions and optical instruments.
Solar plasma transparency for radio emission increases with its
frequency. The atmospheric transparency window enables
ground-based observatories to receive solar radio emission from
tens of MHz to hundreds of GHz and to gain information from
emitting areas from the chromosphere to the high corona. The range
of the time scales under study is also extensive, from millisecond
pulses generated by coherent mechanisms in abnormally bright
compact sources to multi-year solar cycles.

Some physical parameters of the solar corona can only be estimated
from radio observations. The geoeffective potential of eruptions
of solar magnetized plasma, which cause sporadic disturbances in
near-Earth space, depends on coronal magnetic fields; the rate and
intensity of eruptive processes are determined by dynamics of
magnetic energy release and its conversion into energy of plasma
particles and emission. Therefore, one of the most important tasks
is to monitor the evolution of magnetic structures in the solar
coro\-na. The brightness of magnetically active lines in the
corona is too low against the photospheric background that hampers
magnetic field measurements in the corona with optical
instruments. Microwave observations are capable of solving this
problem, because the microwave spectrum and polarization strongly
depend on the strength and orientation of the magnetic field in a
source. Radio magnetography of the solar corona becomes possible
with the advent of new generation radioheliographs which provide
sequences of solar disk images at different frequencies (see, for
example, \citealp{Lang1993, Wang2015}). An important feature of
radio observations is their high sensitivity to emission of
nonthermal electrons. Radio observations are sensitive enough to
nonthermal emission from coronal regions where the plasma density
is too low to produce detectable hard X-rays.

Observations of eruptive processes in the solar corona are
important to improve physical background of forecasting and
diagnostics of geoeffective phenomena such as solar flares and
coronal mass ejections. Radioheliographs detect microwave bursts
associated with energy release processes and reveal their sources
on the solar disk.

Flare processes often produce radio bursts whose intensity is
millions of times higher than background. Data on their spectrum
and sources are important for diagnostics of sporadic solar
activity, which is necessary to analyze causes of malfunctions in
electronic systems on spacecraft, radars, navigation and
communication systems. Generalization and systematization of these
data have important implications for hardware developers.

Objectives of observations in the microwave range involve
identifying predictors of powerful solar flares, excitation of
shock waves, and appearance of accelerated electron and proton
fluxes in interplanetary space.

Since the angular resolution of a telescope depends on the ratio
of received emission wavelength to its aperture, observation of
the evolution of active regions and flare sources requires
developing instruments of hundreds of meters in size. To reach the
above objectives, it is important to combine sufficiently high
spatial resolution with a wide field of view exceeding the solar
disk that is possible to attain with multi-element
interferometers. Their spatial resolution depends on the aperture
of the antenna array; and the field of view is determined by
individual antenna elements. Spectral resolution should be high
enough to identify emission mechanisms, regions with optically
thin and thick emission, characteristics of emitting electrons,
etc.

Important information on emission mechanisms is carried by
polarization, insufficiently used so far to interpret observations
because of its dependence on the spatial structure of sources. The
analysis of polarization data is often complicated by propagation
effects of the emission from the source in surrounding plasma.
These effects cause absorption, polarization reversal, and
scattering. On the other hand, if these effects are significant,
they allow estimating plasma parameters on the way from the source
to an observer.

Until recently, the main instrument of the ISTP SB RAS
Radioastronomical complex was SSRT, which commenced observations
in 1986 \citep{Smolkov1986, Grechnev2003ssrt}. SSRT is included in
Russia's list of unique instruments, and the team of designers got
the Russian Federation Government Award. A large amount of digital
data have been accumulated from daily observations at a frequency
of 5.7\,GHz during daylight hours in the form of two-dimensional
radio maps and one-dimensional distributions of radio brightness.
At present, SSRT is transformed into a new instrument, multiwave
Siberian Radioheliograph \citep{Lesovoi2014}. In 2016, the
first-stage 48-antenna array (SRH-48) commenced observations in a
frequency band of 4--8 GHz.

Previously, concurrent solar imaging was possible at three
frequencies: 5.7\,GHz at SSRT, 17 and 34\,GHz at the Japanese
Nobeyama Radioheliograph (NoRH; \citealp{Nakajima1995}). The
observational daytime interval of NoRH from 23 to 06\,UT has a
large overlap with that of SSRT. There is also an overlap with the
RATAN-600 radio telescope, which measures one-dimensional radio
brightness distribution over the solar disk in a wide frequency
range with a high spectral resolution \citep{Bogod2011,
Kaltman2015}. Multiwave observations provide a wealth of
information on the quiet solar atmosphere and active processes in
it; the joint analysis of microwave observations with data from
different spectral ranges gives a unique insight into various
solar processes, their interrelation and effect on near-Earth
space (see, e.g., \citealp{AltyntsevKashapova2014}). The current
state of digital electronics and computer technology makes it
possible to develop multi-element radio interferometers producing
solar images at a number of frequencies almost simultaneously.

Current implementation of several projects confirms that
new-generation radioheliographs are required. In the micro\-wave
range, the largest radio telescopes (with a baseline of about one
kilometer or more) are the American Frequency Agile Solar
Radiotelescope (FASR; \citealp{Bastian1998}) and the Chinese
Mingantu Ultrawide Spectral Radioheliograph (MU\-SER;
\citealp{Yan2009}). The projects are currently at different stages
of implementation. Under the FASR project, the Expanded Owens
Valley Solar Array (18 antennas, 1--9\,GHz) is employed to devise
and test radioheliograph systems. Efforts are underway to increase
the number of antennas and to expand the receiving band up to
2.5--18\,GHz \citep{Gary2012}. In MUSER, located in Inner
Mongolia, antenna systems of CSRH-I (40 antennas, 0.4--2.0\,GHz)
and CSRH-II (60 antennas, 2--15\,GHz) have been installed. Test
observations at some frequencies have started. Efforts are applied
to phase calibration issues.

This paper presents preliminary results of observations made with
the first stage of SRH with the T-shaped 48-antenna array in the
4--8 GHz frequency range. Single-frequency test observations were
made since early 2016. Since July 2016, SRH-48 has routinely been
observing at five frequencies. During this period, solar activity
was low. This allowed us to assess capabilities of the new
instrument to study faint events, which cannot be detected by
total-flux telescopes.

\section{CHARACTERISTICS OF THE SIBERIAN RADIOHELIOGRAPH
WITH THE 48-ANTENNA ARRAY }

The Siberian Solar Radio Telescope, which is transformed into the
Siberian Radioheliograph, has the following characteristics
\citep{Grechnev2003ssrt}. SSRT is a cross-shaped interferometer
consisting of two EW and SN arrays, each with 128 antennas of
2.5\,m in diameter. The baseline is as long as 622.3\,m, which
determines the angular resolution up to $21^{\prime \prime}$ in
two-dimensional images and up to $15^{\prime \prime}$ in
one-dimensional radio brightness distributions. Both circular
polarized components (RCP and LCP) are measured. The 14\,ms period
of polarization modulation determines the highest temporal
resolution of one-dimensional measurements. In the two-dimensional
mode, images are formed as a result of the change of the Sun's
position relative to interference maxima, firstly, because of the
simultaneous receiving at different frequencies in the
5.67--5.79\,GHz band and, secondly, because of Earth's diurnal
rotation. The latter determines the transit time of the
interference maximum across the Sun; therefore SSRT can produce
images every 2--3 minutes at most.

\begin{figure*} 
\centerline{\includegraphics[width=0.7\textwidth]
    {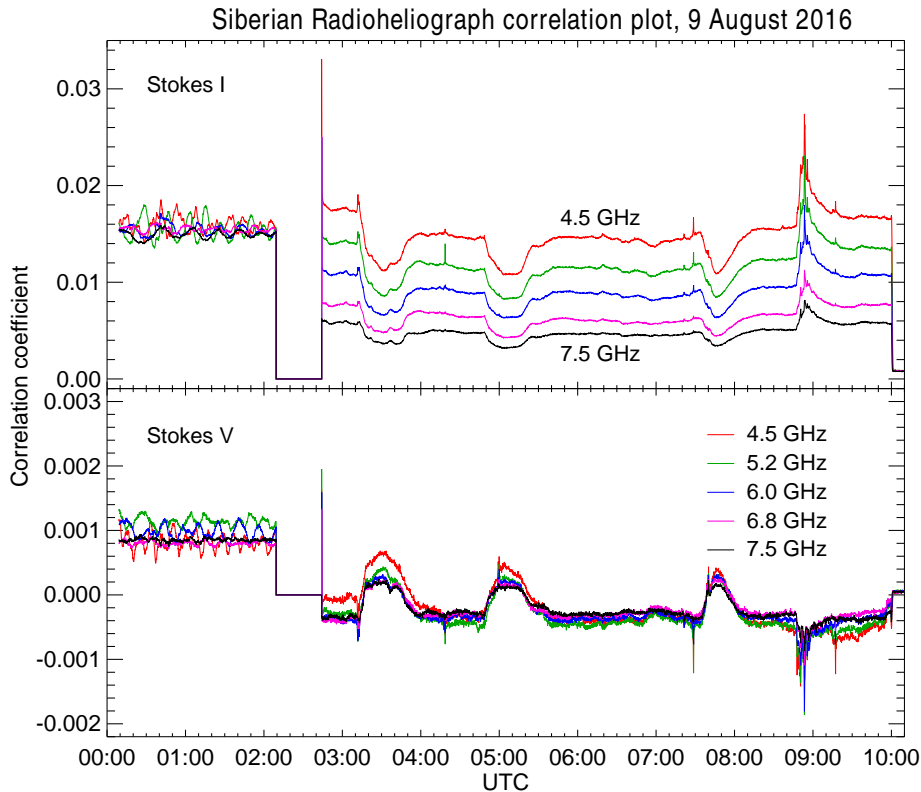}
                      }
\caption{Correlation plots with three negative bursts and a C2.5
flare at 08:55 for 9 August 2016.}
 \label{fig:corr2016-08-09}
\end{figure*}

The Siberian Radioheliograph uses an essentially different imaging
principle, Fourier synthesis \citep{Lesovoi2014}. The temporal
resolution determined by the receiver system is much higher. At
present, routine observations are performed using the 48-antenna
array with antennas installed on the central posts of SSRT along
the east, west, and south arms. Adjustment of SRH-48 systems is
still in progress. The observing frequencies, each of the 10\,MHz
bandwidths in the 4--8\,GHz range, are set by software. The time
to switch from one frequency to another is currently about 2\,s,
and the accumulation time at each frequency is 0.28\,s for each
circularly-polarized component. The maximum baseline is 107.4\,m,
and the spatial resolution is as high as $70^{\prime \prime}$ at
8\,GHz. Both circularly-polarized components are measured. The
sensitivity to compact sources reaches $10^{-4}$ of total solar
flux \citep{LesovoiKobets2017}.

The number of observing frequencies can be changed depending on
the observational program. For quasi-stationary objects such as
sunspot-associated sources, several hundreds of frequencies can be
used to achieve a desired spectral resolution. Observations of
flares require high temporal resolution, for which the number of
frequencies can be reduced. Since 1 July 2016, observations have
been made at five frequencies -- 4.5, 5.2, 6.0, 6.8, and 7.5\,GHz.
Raw data are available at
\url{ftp://badary.iszf.irk.ru/data/srh48}. Software has been
developed and tested to produce raw solar images, clean them, and
calibrate in brightness temperature units. File formats for
storage of data and software for their remote access are being
developed.

Along with observations at SRH-48, the rest of SSRT's antennas
continue observations at 5.7\,GHz in the original operating mode.
All antennas of the north SSRT arm and outer antennas of other
arms provide two-dimensional images of compact structures in the
solar atmosphere with a resolution of up to $21^{\prime \prime}$
and a 2--3\,min interval. Daily maps in intensity and circular
polarization near local noon are available at
\url{ftp://badary.iszf.irk.ru}.

To monitor solar activity and main SRH systems, the so-called
correlation plots are used. They represent a proxy of radio flux
and display temporal variation in the sum of cross-correlations of
all antenna pairs. Methods for calculating the correlation plots
and their relation with characteristics of solar emission are
discussed in \cite{LesovoiKobets2017}. Changes in the correlation
plots are associated with variations both in the brightness of
sources and in their structure. Real-time correlation plots and
quick-look images produced by SRH at five frequencies are
accessible online at \\
\url{http://badary.iszf.irk.ru/srhCorrPlot.php}.

The correlation plot for 9 August 2016 in
Figure~\ref{fig:corr2016-08-09} is interesting due to the presence
of three negative bursts and subsequent C2.5 flare in active
region 12574 (N04E59). Maintenance of SRH systems was carried out
during first hours of that day. Observations started at 02:45\,UT.

\begin{figure*} 
\centerline{\includegraphics[width=\textwidth]
       {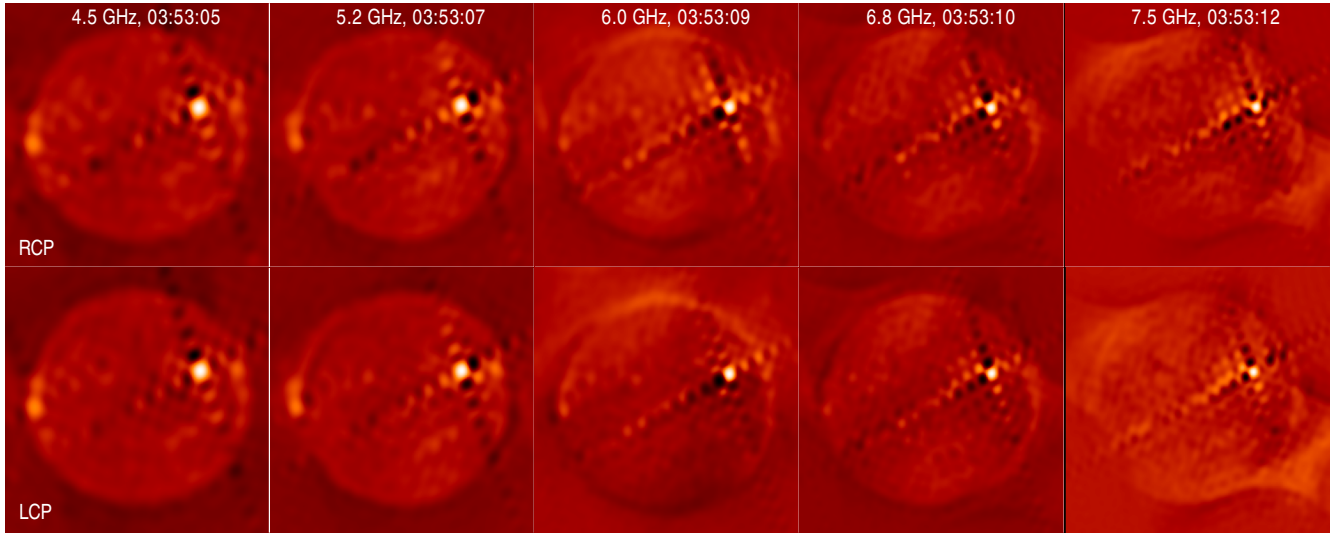}
           }
\caption{Images of the Sun at five frequencies on 9 June 2016 in
the right (RCP) and left (LCP) circularly polarized emissions.}
 \label{fig:radiomap}
\end{figure*}

Figure~\ref{fig:radiomap} exemplifies a set of solar images
observed at five frequencies in the right-handedly and
left-handedly polarized emission without cleaning the images from
the contribution of side lobes of the interferometer. A bright
source on the north-west is a response to active region
NOAA\,12571. Side lobes look like narrow zebra stripes. The images
at frequencies below 6\,GHz also show a brightening on the east
limb, which is associated with rising active regions. The field of
view of the images is one-third larger than the solar disk. The
apparent size of a bright compact source decreases with frequency,
being determined by convolution of the real source with the
frequency-dependent interferometer beam.

\begin{figure*} 
 \centerline{\includegraphics[width=0.75\textwidth]
    {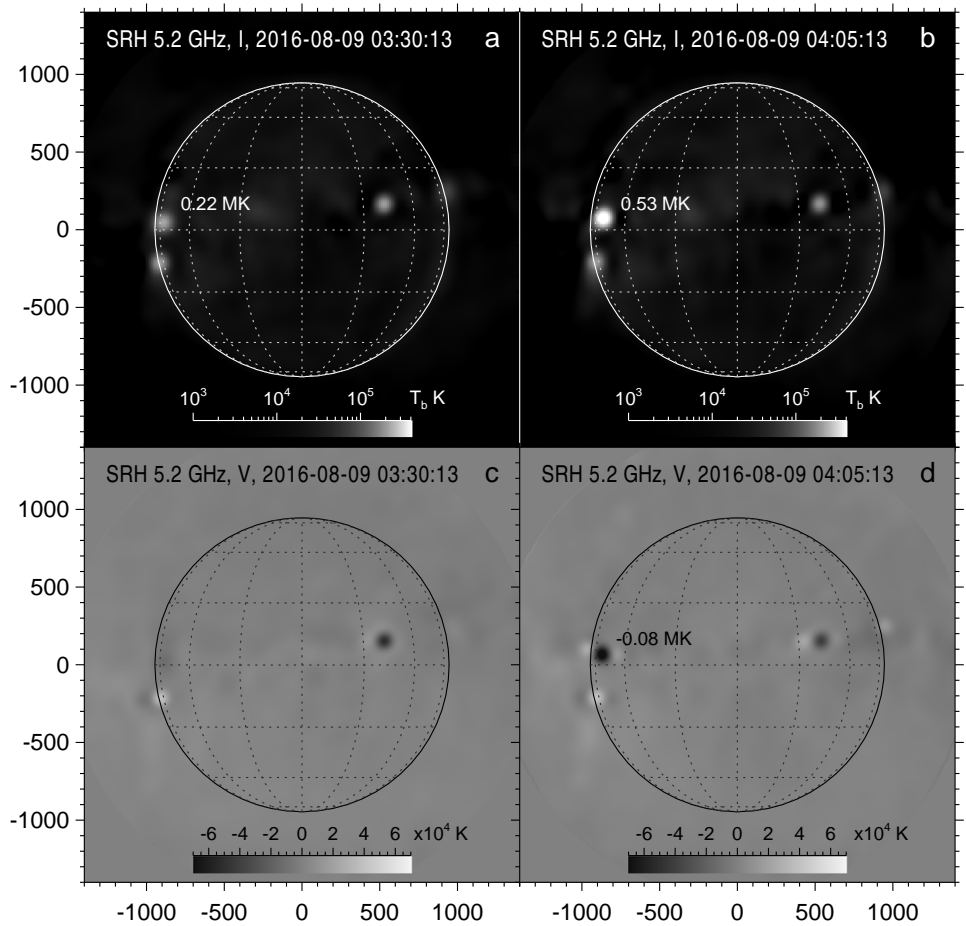}
                  }
\caption{Solar radio images obtained by SRH at 5.2\,GHz in total
intensity and polarization after cleaning procedure during the
first negative burst (left) and after it (right). The negative
burst is related to the screening of the northern source on the
east limb. The axes show arc seconds from the solar disk center.}
 \label{fig:screened_source}
\end{figure*}

To calibrate images in brightness temperatures, the technique
employed at NoRH and SSRT has been elaborated. For each image,
histograms of the brightness distributions are computed for the
regions occupied by the sky and the solar disk. The difference
between the positions of the histograms' peaks represents the
brightness temperature of the quiet Sun \citep{Kochanov2013}. This
temperature, decreasing with increasing frequency, is referred to
the measurements made in \cite{Zirin1991} and \cite{Borovik1994}.
In particular, at the five SRH frequencies, the values of 18.7,
17.1, 15.4, 14.3, and 13.5 thousand Kelvin are adopted. Brightness
temperatures of compact sources can be considerably lower than the
real temperatures if the source is less than the beam pattern of
SRH-48.

\section{PRELIMINARY RESULTS OF FIRST OBSERVATIONS}

Temporary depressions of radio flux below the quasistationary
level (negative bursts) sometimes occur. They are caused by
screening of emission from compact radio sources or quiet solar
regions in low-temperature plasma ejected into the solar corona
during eruptions. The dependence of the absorption depth on the
emission frequency and properties of absorbing plasma allows
estimating the temperature, density, and size of the eruptive
structure if a negative burst is recorded at several radio
frequencies \citep{Grechnev2013}. Negative bursts provide valuable
information about eruptive events, but they are very rarely
observed. The maximum number of negative bursts recorded by all
ground-based stations during a year was as small as 14 in 1991.
The occurrence of three negative bursts during one day on August
9, 2016 is unprecedented (Figure~\ref{fig:corr2016-08-09}).

SRH images in Figure~\ref{fig:screened_source} show that these
negative bursts were caused by a decrease in the brightness of the
northern source near the east limb. Its maximum brightness
temperature near the minimum of the first negative burst in
Figure~\ref{fig:screened_source}a was 0.22\,MK. After the end of
the negative burst, the brightness temperature of the screened
source increased to 0.53\,MK (Figure~\ref{fig:screened_source}b),
whereas in the other three sources its change did not exceed
13\,\%. The last value is the upper boundary of the total error in
measurements of brightness temperatures from SRH data caused by
both the calibration instability and flat-field issues. The
brightness temperatures in the images are lower than those
typically observed because of insufficient spatial resolution of
the current configuration of SRH-48.

\begin{figure*} 
 \centerline{\includegraphics[width=0.7\textwidth]
     {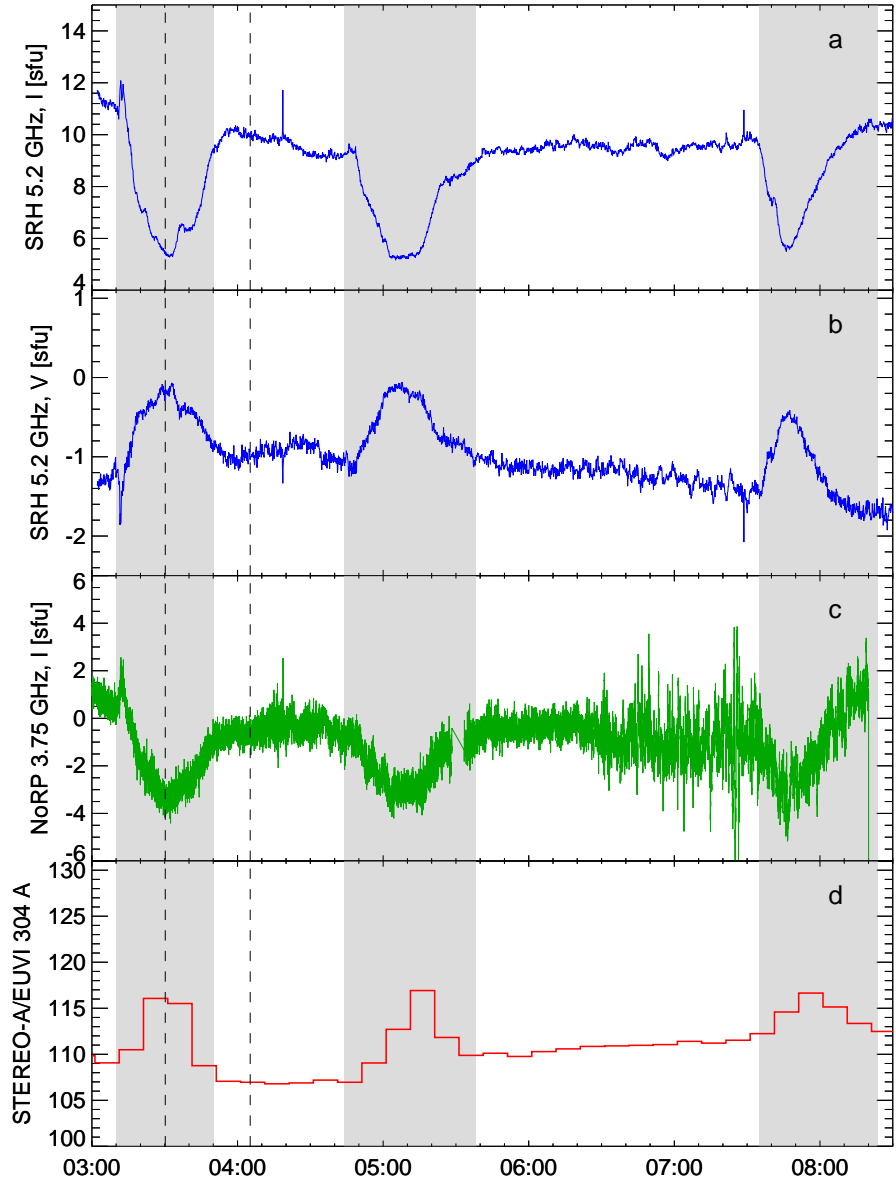}
           }
\caption{Observations on 9 August 2016. Time profiles in total
intensity (a) and polarization (b) of the screened source computed
from SRH images at 5.2\,GHz along with a total flux recorded by
NoRP at 3.75\,GHz (c) and the ultraviolet flux in 304\,\AA\ for
the framed region in Figure~\ref{fig:surge}d. Vertical dashed
lines indicate the observation times in
Figure~\ref{fig:screened_source}. Gray vertical shadings denote
the intervals to which images in Figure~\ref{fig:surge}
correspond.}
 \label{fig:neg_bursts}
\end{figure*}

Figure~\ref{fig:neg_bursts} presents total flux time profiles of
the screened microwave source in total intensity and polarization
computed from 2537 pairs of images observed with an interval of
8.4\,s. The comparison with the total flux at 3.75\,GHz, recorded
by NoRP with an interval of 1\,s, demonstrates the solar origin of
the negative bursts and high sensitivity of SRH. With intensity
depressions reaching $-5$\,sfu (1\,sfu = $10^{-22}$ W/(m$^2$\,Hz))
at 5.2\,GHz the polarization shifted to positive values. This
corresponds to screening of the left-polarized source
(Figure~\ref{fig:screened_source}c,\,d).

\begin{figure*} 
 \centerline{\includegraphics[width=0.8\textwidth]
    {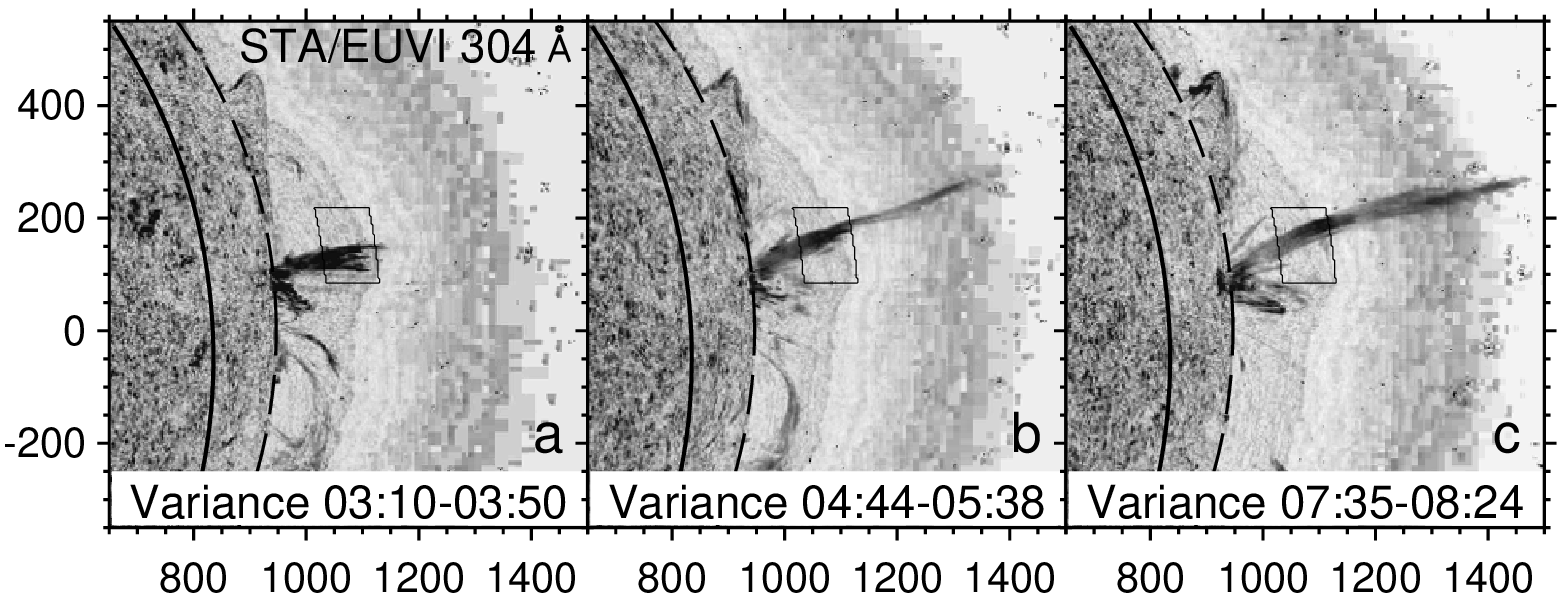}
                  }
\caption{Observations on 9 August 2016. Images of three surges
revealed by the variance analysis of STEREO-A/EUVI 304\,\AA\
images in specified intervals. The frame at the center of the
image indicates the region used to construct the time profiles in
Figure~\ref{fig:neg_bursts}d. The dashed arc denotes the solar
limb visible from STEREO-A. The thick solid arc corresponds to the
east limb visible from Earth. The axes show arc seconds from the
center of the solar disk as seen from STEREO-A.}
 \label{fig:surge}
\end{figure*}

No coronal mass ejections were associated with these negative
bursts and no solar observations were made by SDO/AIA on that day.
Rising jets (surges) were found in 304\,\AA\ images produced by
the STEREO-A space observatory located in the Earth orbit to the
east of it by $152^{\circ}$. The images in Figure~\ref{fig:surge}
are negatives of variance maps revealing all changes in the
images. Each pixel of such a map represents variance of values at
this point in all images taken within a certain time interval
\citep{Grechnev2003vm}. The frame denotes the region whose flux
variations are shown in Figure~\ref{fig:neg_bursts}d. The
occurrence of negative bursts is certainly related to the surges
screening the source. Further analysis of these negative bursts
promises estimating parameters of surges.

\begin{figure*} 
\centerline{\includegraphics[width=0.8\textwidth]
    {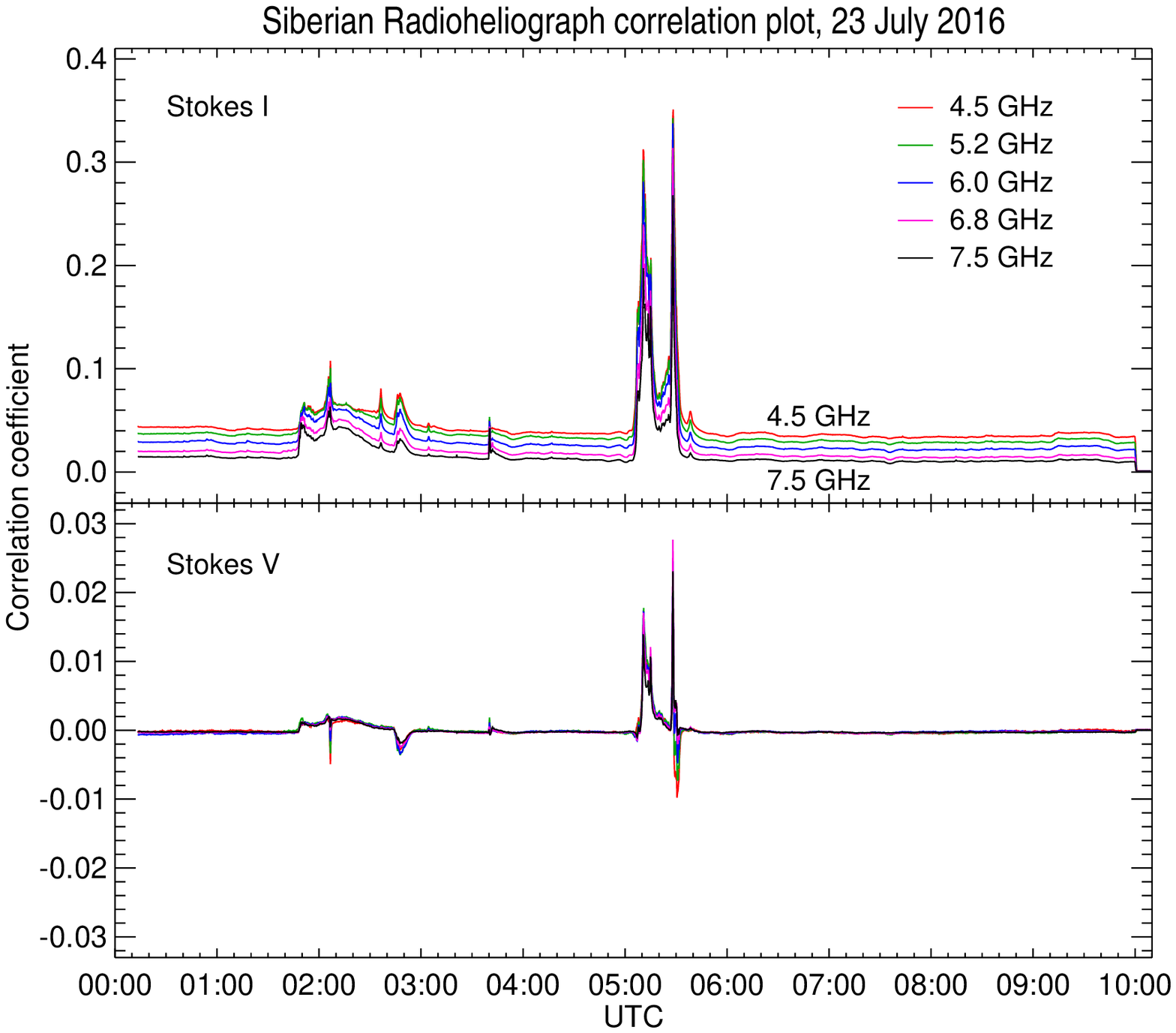}
                  }
\caption{Correlation plots with powerful flares recorded on 23
July 2016.}
 \label{fig:flare_corplot}
\end{figure*}

\begin{figure*} 
\centerline{\includegraphics[width=0.8\textwidth]
    {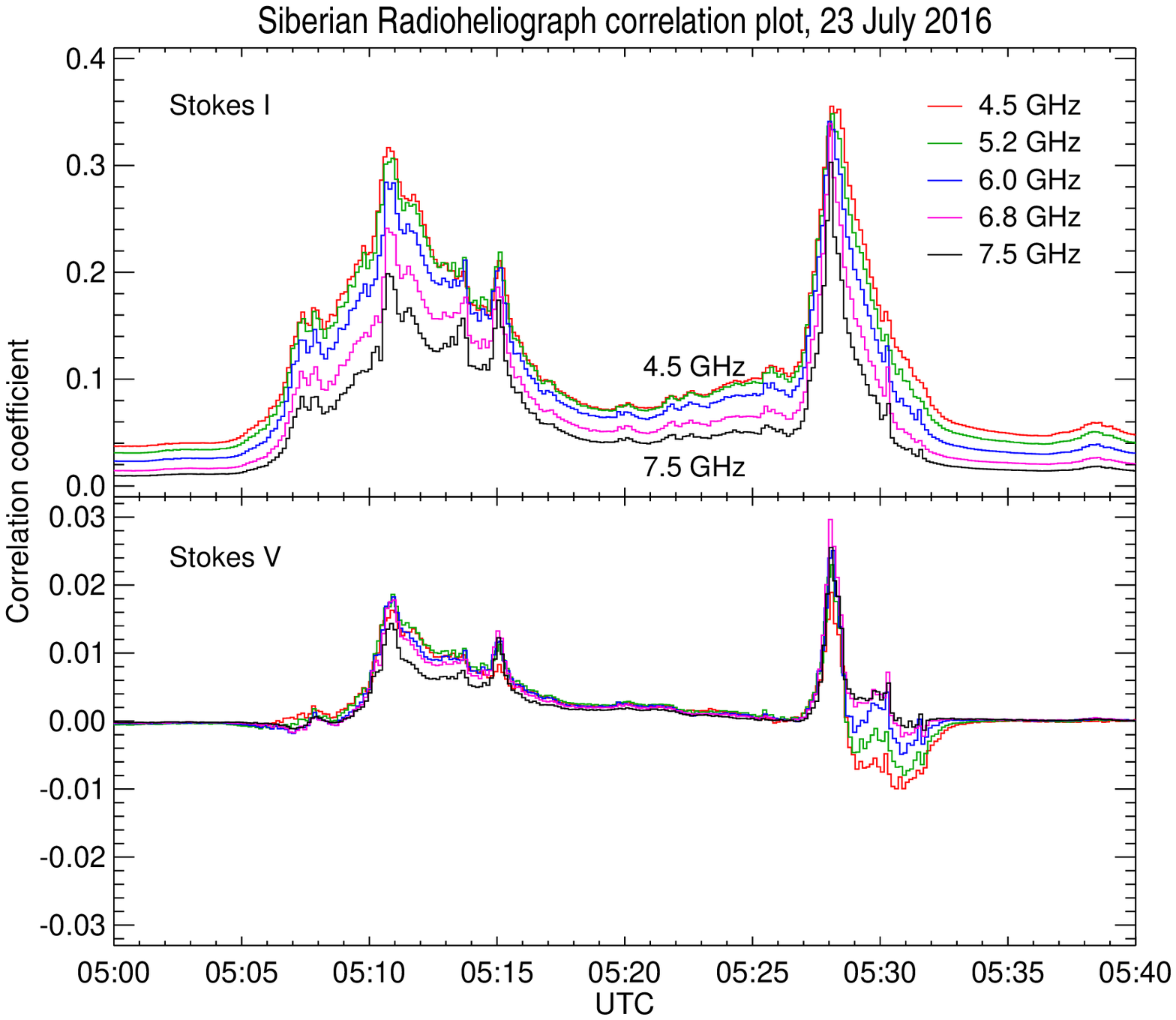}
                  }
\caption{Correlation plots of two flares on 23 July 2016 with
polarization reversal within the SRH frequency band
(05:28:30--05:30:30\,UT).}
 \label{fig:flare_corplot_mag}
\end{figure*}

Along with a high sensitivity, the SRH receiving system has a
sufficiently wide dynamic range to observe the microwave emission
from flares without attenuators. Figure~\ref{fig:flare_corplot}
shows correlation plots with three powerful M-class flares: M5.0
(02:11), M7.6 (05:16), and M5.5 (05:31\,UT) recorded on 23 July
2016. The most intense was the microwave burst in the last flare
whose flux at SRH frequencies exceeded 800\,sfu. Between 05:28:30
and 05:30:30\,UT, polarization reversal occurred within the SRH
frequency range (Figure~\ref{fig:flare_corplot_mag}). The
polarization reversal of flare microwave emission can be caused by
different reasons. To determine them, further analysis of
spatially-resolved data is required.

\begin{figure*} 
\centerline{\includegraphics[width=0.7\textwidth]
     {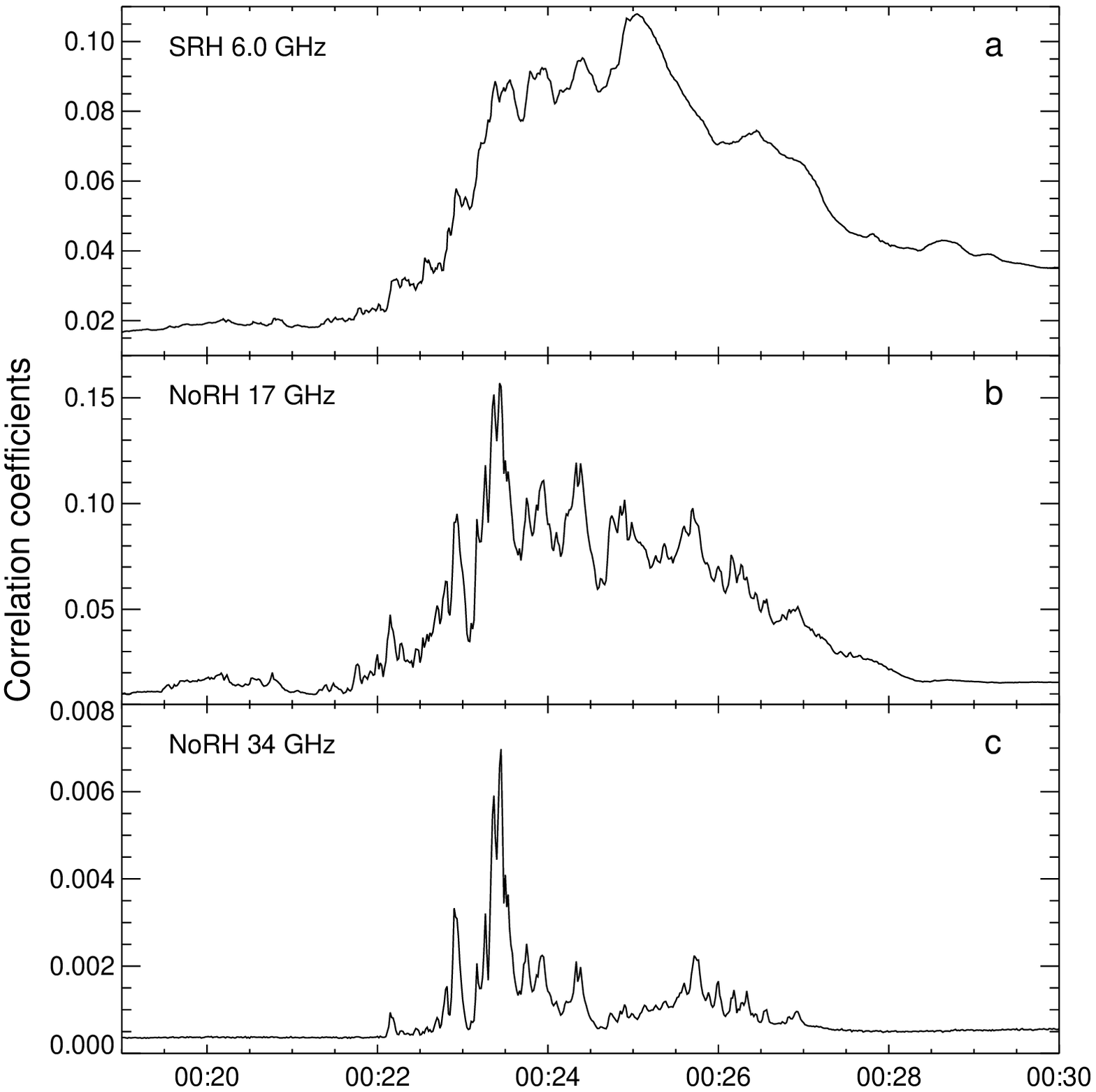}
           }
\caption{Correlation plots of the 18 April 2016 M7.6 flare
recorded by SRH at 6\,GHz (a) and NoRH at 17\,GHz (b) and 34\,GHz
(c).}
 \label{fig:flare_timeprof}
\end{figure*}

Figure~\ref{fig:flare_timeprof} presents correlation plots of the
18 April 2016 M7.6 flare. With a general correspondence between
the SRH (Figure~\ref{fig:flare_timeprof}a) and NoRH
(Figure~\ref{fig:flare_timeprof}b,\,c) data, the burst time
profile, as usual, becomes sharper with increasing frequency. This
event is interesting due to pronounced broadband microwave
pulsations with a period of about 30\,s.

As a rule, the 4--8 GHz receiving band of SRH does not cover the
entire spectrum of gyrosynchrotron emission of powerful solar
flares, which may extend to several tens of GHz. Nevertheless, SRH
images allow identifying important flare structures, for example,
high flare loops invisible at high frequencies
\citep{Altyntsev2016, Fleishman2016}. In most weak to moderate
microwave bursts, the spectral peak falls within the SRH frequency
band or lies close to it \citep{Nita2004}.

A joint analysis of SRH observations with data obtained in
different spectral ranges is particularly promising in studies of
complex solar phenomena. As an example, we show a preliminary
analysis of the 16 March 2016 C2.2 eruptive flare. Before this
event, maintenance of SRH systems was carried out. Antennas were
pointed at the Sun at about 06:36. SRH observed this event at
6\,GHz with an interval of about 1\,s between the images.

\begin{figure*} 
\centerline{\includegraphics[width=\textwidth]
    {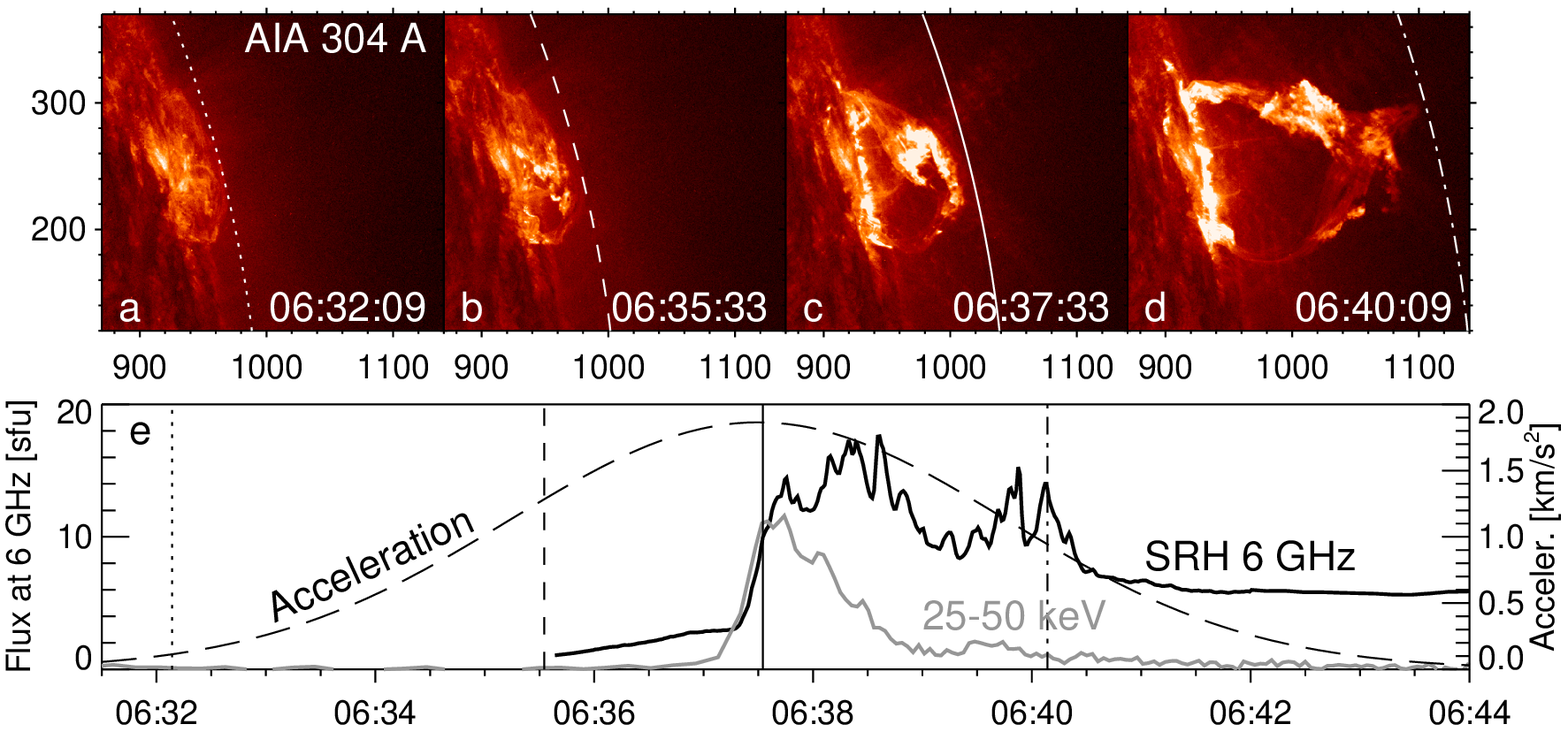}
           }
\caption{Eruptive flare of 16 March 2016. Top: prominence eruption
in SDO/AIA 304\,\AA\ images. The axes show arcseconds from the
solar disk center. Arcs outline the upper edge of the rising
prominence. Bottom: time profiles of the microwave (black, SRH
6\,GHz) and hard X-ray (gray, Fermi 25--50\,keV) emissions along
with the measured acceleration of the prominence (dashed curve)
corresponding to the arcs on panels a--d. Times of these images
are marked on the bottom panel by vertical lines of the same
styles as the arcs.}
 \label{fig:eruption}
\end{figure*}

Figure~\ref{fig:eruption} presents a prominence eruption near the
west limb, which was observed in the 304\,\AA\ channel of the
Atmospheric Imaging Assembly (AIA; \citealp{Lemen2012}) of the
SDO. Arcs of different styles in Figure~\ref{fig:eruption}a--d
outline the upper edge of the ascending prominence, and the dashed
curve in Figure~\ref{fig:eruption}e represents the fit of its
acceleration, which is up to 1.8\,km\,s$^{-2}$.

For comparison, Figure~\ref{fig:eruption}e also presents time
profiles of hard X-rays recorded by the Fermi Gamma-Ray Burst
Monitor \citep{Meegan2009} and of the microwave burst computed
from SRH images at 6.0\,GHz. The prominence started to ascend a
few minutes before the sharp onset of the burst in hard X-rays and
microwaves. Acceleration of most electrons up to high energies in
the flare, which are responsible for the hard X-ray and microwave
burst, was clearly caused by the prominence eruption. This
sequence of events is typical of eruptive flares
\citep{Grechnev2015, Grechnev2016}.

\begin{figure*} 
\centerline{\includegraphics[width=0.72\textwidth]
    {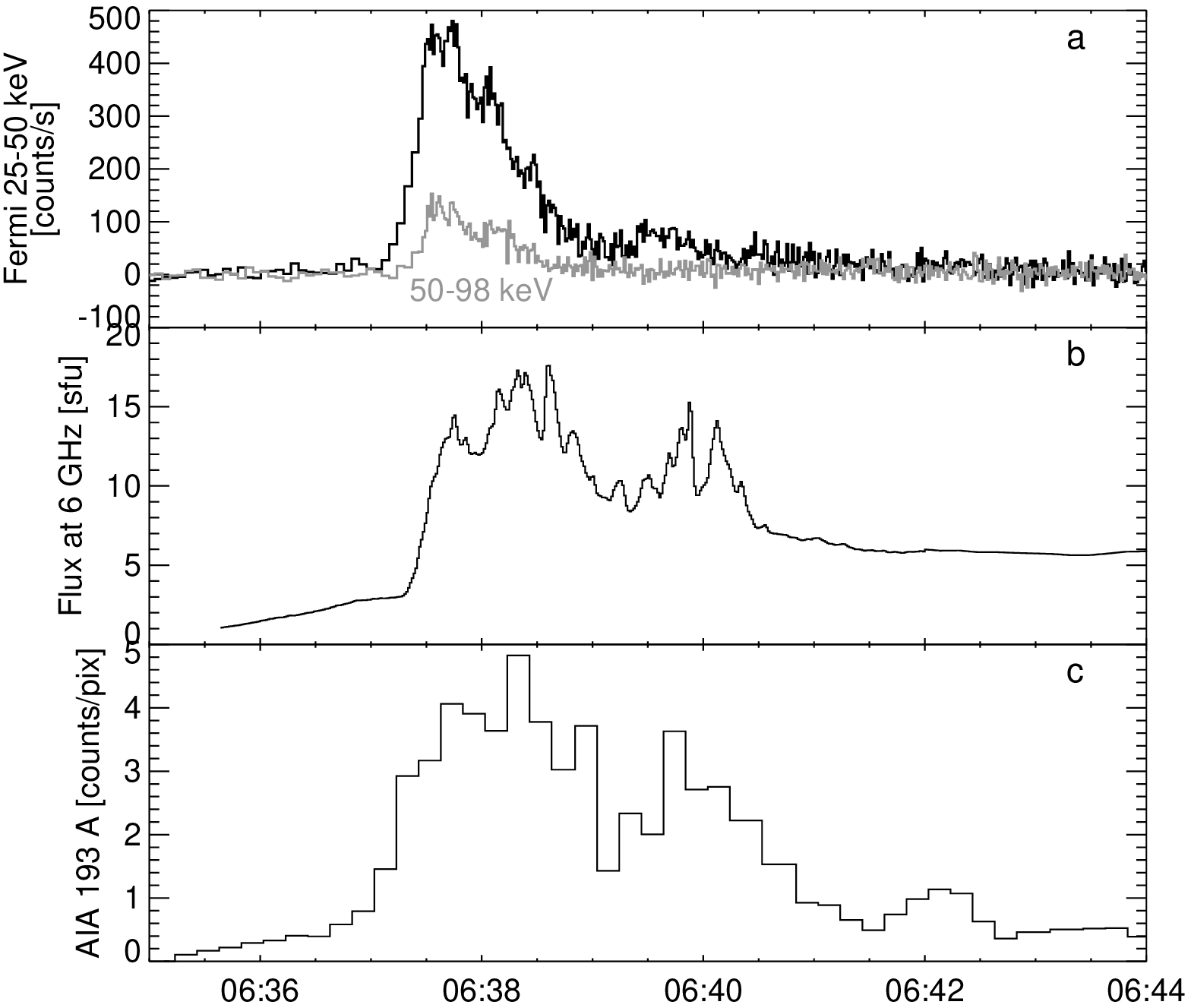}
           }
\caption{Time profiles of hard X-ray (Fermi/GBM, a), microwave
(SRH, b), and extreme-ultraviolet (SDO/AIA 193\,\AA, c) emissions
for the 16 March 2016 flare. Each point of the time profile (c)
was calculated from the difference between the current SDO/AIA
193\,\AA\ image and the image observed 48\,s earlier within bright
regions above the flare ribbons.}
 \label{fig:srh_hxr_euv}
\end{figure*}

\begin{figure*} 
\centerline{\includegraphics[width=0.75\textwidth]
    {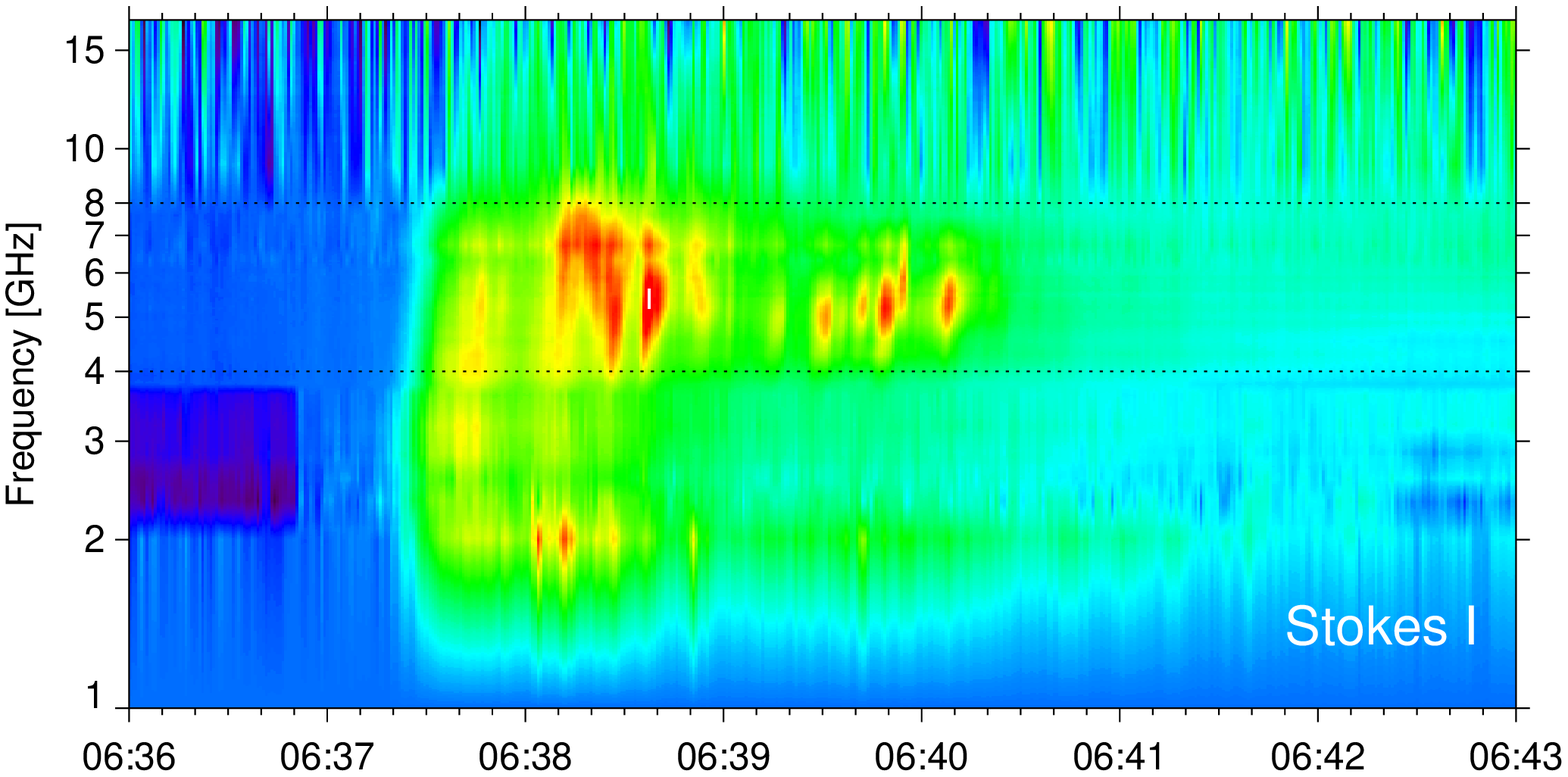}
           }
\caption{Dynamic spectrum of the 16 March 2016 flare. Horizontal
dotted lines indicate the SRH frequency band.}
 \label{fig:spectrum}
\end{figure*}

Figure~\ref{fig:srh_hxr_euv} shows time profiles of the flare in
different spectral ranges: in two hard X-ray channels
(Fermi/GBM,~a), in microwaves at 6 GHz (SRH,~b), and in extreme
ultraviolet (SDO/AIA 193\,\AA, c). The last time profile is
calculated from running difference images (the image observed
48\,s before each current image is subtracted from it) within
bright regions above flare ribbons $\leq 4^{\prime \prime}$  wide
in the plane of the sky, i.e. at a height up to 3000\,km (these
regions are also seen in 304\,\AA\ images in
Figure~\ref{fig:eruption}a,\,d). The temperature sensitivity
region of the 193~\AA\ channel starts from 0.2\,MK ($ > 2 \cdot
10^{5}$\,K). Consequently, the time profile in
Figure~\ref{fig:srh_hxr_euv}c represents hot dense regions above
the flare ribbons whose prolonged emission is suppressed in
running differences. The similarity between the time profiles at
6\,GHz and in the 193\,\AA\ channel leaves no doubt that they have
common sources. There is no such similarity with the time profiles
in the 193\,\AA\ channel for each of the separate regions above
the flare ribbons. Hence, the sources of microwave emission were
distributed throughout the whole lengths of the ribbons, being
most likely located in the lower parts of the flare arcade.

The similarity of the microwave burst with hard X-rays in
Figure~\ref{fig:srh_hxr_euv}a is less pronounced than with the
time profile in the 193\,\AA\ channel. The rapid emission increase
at 6\,GHz occurred almost simultaneously with the hard X-ray
increase, but a slow rise in microwave emission had started
earlier. The initial part of the observations at 6\,GHz is less
reliable because of inaccurate SRH antenna pointing at the onset
of the event; yet a similar emission increase in the 193\,\AA\
channel confirms correctness of the SRH data. The second peak
after 06:39 is also identified in an emission band 25--50 keV,
though being weaker than in microwaves, but it is not seen in
harder X-rays 50--98 keV. This suggests that electrons in the
second peak have a softer spectrum. The long-term background of
the microwave burst is most likely the sum of thermal
bremsstrahlung from plasma in the flare arcade and emission of
accelerated electrons trapped in its loops. Despite the
differences in microwave and hard X-ray emissions, these bursts
and their individual structural details bear general similarity.
The sources of hard X-rays were most probably distributed over the
length of the flare ribbons too. This agrees with widely accepted
model predictions.

A combined microwave spectrum of this event in
Figure~\ref{fig:spectrum} inferred from data acquired by
total-flux spectropolarimeters \citep{ZhdanovZandanov2015} and by
NoRP shows a series of 5--10\,s pulses, whose spectral maximum
frequencies were $<10$\,GHz. The spectrum widths of these
impulsive bursts during the second peak (06:39:00--06:40:30\,UT)
are within 2--3\,GHz, being atypically narrow for gyrosynchrotron
emission even for the soft spectrum of radiating electrons. The
differences between the spectra of microwave pulses suggest the
difference between their sources, thus confirming the conclusion
about their location in different places above the flare ribbons.

Another feature of this flare is a modest microwave flux ($\leq
18$\,sfu) with a rather y intense burst in hard X-rays. One of the
causes of the low microwave emission could be compactness of its
sources.

The preliminary analysis of the 16 March 2016 eruptive flare
observed by SRH-48 shows its capabilities even with incomplete
antenna array reducing its current spatial resolution. The joint
analysis of the SRH and spectropolarimeter data with observations
in different spectral ranges provides insight into the
relationship between eruptions and flares and into properties of
flare configurations even though their being unresolved.
Observations made using instruments with limited spatial
resolution and dynamic range suggest that non-thermal processes
predominate in simple one- or two-loop flare configurations. This
is difficult to reconcile with observations of emission of other
types (e.g., in extreme ultraviolet) and with well-known models.
The correspondence between various flare manifestations observed
in different ranges in the event overviewed is in agreement with
models and conclusions of recent studies (e.g.,
\citealp{Grechnev2017}).

\section{CONCLUSION}

Observations made using the first stage of the multiwave
Sib\-erian Radioheliograph demonstrate efficiency of design ideas
and their accomplishment. The T-shaped antenna array with
redundant baselines enables the implementation of fast and
effective algorithms for solar imaging without the need of
reference observations of other cosmic sources. The high
sensitivity of the interferometer ($\approx 10^{-2}$\,sfu)
combined with a wide dynamic range allows observing compact
sources of powerful solar bursts without attenuators. Advantages
of SRH are as follows: the temporal resolution high enough to
study many processes (up to 0.56\,s for both circularly-polarized
components in the single-frequency mode), multi-frequency
observations with a tunable frequency set depending on
observational program, image synthesis with optimization of
required parameters (for example, spatial resolution or
sensitivity), and the absence of geometric distortions from which
SSRT images suffered.

SRH allows one to synthesize tens of thousands of solar images per
day. Methods and software for synthesis and calibration of images
and their subsequent analysis have been partially developed, in
particular in our previous studies. They should be elaborated and
adapted to address diverse observational and research issues.

Expansion of the antenna array up to 96 elements would be the next
milestone in upgrading SRH. The spatial resolution of SRH-96 would
be as high as $15^{\prime \prime}$. Additional antennas have
already been installed, in particular in outer SSRT posts, which
are located 330\,m away from the center of the array. As SSRT
observations showed, such resolution would be sufficient to study
processes of initiation of coronal mass ejections and their
propagation up to heights of one to two solar radii, thus filling
the gap between observations in ultraviolet and optical ranges
\citep{Uralov2002, Alissandrakis2013}. Multi-frequency
observations of active regions would provide important
verification of coronal magnetography methods \citep{Nita2011}.

The first results presented here show a high potential of the new
instrument, being related to only a few observational issues. The
preliminary analysis of the events discussed needs to be
completed. Some mentioned observational results require
understanding. At the same time, ongoing SRH observations provide
new information. Any interest in the development of the SRH
software and data analysis is welcome.

\phantomsection
\section*{Acknowledgments}
We are grateful to P.M. Svidsky for useful discussions. This work
was supported by Fundamental Research Program~7 of the RAS
Presidium ``Experimental and theoretical studies of objects in the
solar system and planetary systems of stars. Transitional and
explosive processes in astrophysics'', by RFBR under grants
15-02-01089\,A and 15-02-03717\,A, and by FASO under the project
``Investigation into extremely weak solar activity in the
microwave range'' performed at SSRT. Experimental data were
acquired using SSRT. The authors thank the editorial team for the
assistance in polishing the manuscript and preparing its English
version.

\end{document}